\documentclass[aps,twocolumn,pre]{revtex4-1}
\usepackage[T1]{fontenc}
\usepackage{amsmath,amsfonts,amsthm,amssymb}
\usepackage{graphicx}                          
\usepackage{caption}
\usepackage{listings}
\usepackage{tabularx}
\usepackage{xcolor}
\usepackage{indentfirst}
\usepackage{subcaption}
\usepackage{overpic}

\usepackage{enumerate}

\usepackage{bm}

\usepackage[font=footnotesize,labelfont=bf]{caption}

\newtheorem{remark}{Remark}[section]

\newcommand\dd{\mathrm{d}}
\newcommand\pp{\partial}

\newcommand\x{\bm{x}}
\newcommand\uvec{\mathbf{u}}

\newcommand\RT{\mathsf{RT}} 

\usepackage{mhchem}

\begin{document}



\title{Field Theory of Reaction-Diffusion: Law of Mass Action with an Energetic Variational Approach}



\author{Yiwei Wang}
\email{ywang487@iit.edu}
\author{Chun Liu}%
 \email{cliu124@iit.edu}
\affiliation{Department of Applied Mathematics, Illinois Institute of Technology, Chicago, IL 60616, USA.
}%


\author{Pei Liu}
\email{liu01304@umn.edu}
\affiliation{School of Mathematics, University of Minnesota, Minneapolis, MN 55455, USA.
}%

\author{Bob Eisenberg}
\email{beisenbe@rush.edu}
\affiliation{Department of Applied Mathematics, Illinois Institute of Technology, Chicago, IL 60616, USA.
}%
\affiliation{Department of Physiology and Biophysics, Rush University, 1750 W. Harrison, Chicago IL 60612.
}

%

\begin{abstract}
  We extend the energetic variational approach so it can be applied to a chemical reaction system with general mass action kinetics.
  Our approach starts with an energy-dissipation law. We show that the chemical equilibrium is determined by the choice of the free energy and the dynamics of the chemical reaction is determined by the choice of the dissipation. This approach enables us to couple chemical reactions with other effects, such as diffusion and drift in an electric field. As an illustration, we apply our approach to a non-equilibrium reaction-diffusion system in a specific but canonical setup. We show by numerical simulations that the input-output relation of such a system depends on the choice of the dissipation.
\end{abstract}

\maketitle

%

%

\section{Introduction}




Many biological processes can be abstractly represented as biochemical networks, in which chemical reactions are catalyzed by enzymes and combined to perform many of the functions of life. Examples include metabolic pathways and the electron transport chain that power life 
\cite{alon2003biological, okada2016law}. In these systems, 
reactions occur in different physical locations, so the products of one reaction move, by diffusion (and perhaps migration and convection), to become reactants for another reaction 
in a different location. To describe a complex biological system, 
and consistently deal with the coupling between reaction and diffusion, as well as other mechanical effects,  one needs to turn to a variational theory. The variational principle guarantees a consistent mathematical formulation, in which all variables satisfy all equations, of all fields, and their boundary conditions, with one set of parameters in a certain region. 


For mechanical systems, inspired by the seminal work of Rayleigh \cite{strutt1871some} and Onsager \cite{onsager1931reciprocal, onsager1931reciprocal2}, various variational theories have been developed. Examples includes the Energetic Variational Approach (EnVarA) \cite{liu2009introduction, giga2017variational}, the  general equation for the nonequilibrium reversible-irreversible coupling (GENERIC) \cite{grmela1997dynamics, ottinger1997dynamics, grmela2018generic, pavelka2018multiscale}, Doi's Onsager principle \cite{doi2011onsager, doi2015onsager}, and the Conservation-Dissipation Formalism (CDF) \cite{yong2012conservation, peng2019conservation}. However, it is not straightforward to apply these variational principles to a chemical reaction system, which cannot be understood from Newtonian mechanics \cite{ge2016mesoscopic}.

The macroscopic dynamics of chemical reactions are often described by
\emph{the law of mass action}, which states that the rate of a reaction is proportional to the concentrations of the reactants \cite{chang1998chemistry, keener1998mathematical}. 
The law of mass action originally arises from the treatment of ideal gases (solutions) \cite{waage1986studies}, where molecules/atoms only interact when they collide.
Although the mass-action type kinetics has been widely used for different chemical reaction systems, it is a phenomenological theory, of which the underlying physical foundation is unclear, as molecules can interact in many different ways. As aptly pointed out in Ref. \cite{keener1998mathematical}, \emph{``the law of mass action is not a law in the sense that it is inviolable, but rather is a useful model, much like Ohm's law or Newton's law of cooling.''}
Since 1950's,  there has been a huge amount of work devoted to studying the thermodynamics basis and mathematical structures of chemically reacting systems
\cite{wei1962axiomatic, shear1967analog, shapiro1965mass, aris1965prolegomena, sellers1967stoichiometric, aris1968prolegomena, horn1972general, oster1974chemical, perelson1974chemical, kurtz1972relationship, othmer1976nonuniqueness, Bruce1980, biot1982thermodynamic, Truesdell1984, reti1984analogies, beris1994thermodynamics, gorban2004constructive, mielke2011gradient, grmela2012fluctuations, keizer2012statistical, yong2012conservation, liero2013gradient, mielke2013thermomechanical, van2013mathematical, klika2013coupling, gorban2015generalized, anderson2015stochastic, anderson2015lyapunov, rao2016nonequilibrium, qian2016entropy, ge2016mesoscopic, ge2017mathematical, mielke2017non, haskovec2018decay, feinberg2019foundations,fang2019lyapunov, gorban2019new}. In particular, many papers extend the 
variational principle for mechanical systems to the reaction kinetics by building the 
analogies between Newtonian mechanics and chemical reactions \cite{biot1977variational,reti1984analogies, beris1994thermodynamics, kondepudi2014modern}. 

The goal of this paper is to extend the framework of EnVarA \cite{liu2009introduction, eisenberg2010energy, giga2017variational}, which has dealt with flows in systems with many components successfully for many years, to a chemical reaction system. 
We model a reaction system by a prescribed energy-dissipation law, in which the choice of the free energy determines the chemical equilibrium (if it exists), and the choice of the dissipation determines 
the dynamics of the chemical reaction.
The classical law of mass action can be derived from a particular choice of the energy-dissipation law. 
Our approach is non-equilibrium and provides a basis to couple the chemical reaction with the effect of other fields, such as diffusion, drift in an electric field, as well as the thermal effects. 
 As an illustration, we apply our approach to a non-equilibrium reaction-diffusion system, which can be viewed as an abstract building block of biological networks with inputs and outputs. Our analysis shows that the input-output relation of such a system depends on the choice of the dissipation, which indicates the dissipation can be inferred from experimental measurements.

\section{Field Theory of Reaction-Diffusion}\label{set:EnVarA}




To clarify our ideas, we first consider a single reversible reaction 
\begin{equation}\label{rev_reaction}
\ce{\alpha A + \beta B <=> \gamma C},
\end{equation}
where $\alpha$, $\beta$ and $\gamma$ are stoichiometric coefficients. 
Let $c_i$ ($i = A, B, C$) denote the concentration of each species. Due to conservation of elements, 
 one must have 
\begin{equation}\label{constraint}
  \begin{aligned}
    &  \frac{\dd}{\dd t} \left( \gamma c_A + \alpha c_C \right) = 0, \quad \frac{\dd}{\dd t} \left( \gamma c_B + \beta c_C \right) =  0, \\
  \end{aligned}
  \end{equation}
  which is known as the stoichiometric constraint \cite{oster1974chemical}.
As a consequence, 
each component of $\bm c = (c_A, c_B, c_C)^{\rm T}$ satisfies the ordinary differential equation
\begin{equation}\label{RK0}
  \frac{\dd}{\dd t} c_i = \sigma_i r(\bm{c}),
\end{equation}
where ${\bm \sigma} = (-\alpha, -\beta, \gamma)^{\rm T}$ is the stoichiometric vector, and $r(\bm{c})$ is known as the reaction rate \cite{chang1998chemistry}. 

The law of mass action 
gives a particular form of the reaction rate $r({\bm c})$,
\begin{equation}
r({\bm c}) = k_f c_A^{\alpha} c_B^{\beta} - k_r c_C^{\gamma},
\end{equation}
where $k_f$ and $k_r$ are rate constants for forward and reverse directions.
This form originally arose from the treatment of ideal gases (solutions) \cite{waage1986studies}, where molecules/atoms only interact when they collide.
At an equilibrium, in which the concentrations are not changing, we have
\begin{equation}
\frac{(c_A^{\infty})^{\alpha} (c_B^{\infty})^{\beta}}{(c_C^{\infty})^{\gamma}} =  \frac{k_r}{k_f} \triangleq K_{eq},
\end{equation}
where $K_{eq} = \frac{k_r}{k_f}$ is called the equilibrium constant. The law of mass action 
is an empirical law without a clear physical interpretation, so it can not be immediately applied to biological and electrochemical systems in which chemical reactions are coupled with other effects.

In the remainder of this section, we show that the simple reaction kinetics (\ref{RK0}) can be modeled by an extended energetic variational approach,
like a mechanical system, which provides a basis of coupling chemical reaction with other mechanisms, including mechanical effects such as diffusion, drift in an electric field, as well as the thermal effects. As an application, we also provide an energetic variational formulation to a reaction-diffusion system, which is a typical example of mechano-chemical or chemomechanical systems.

\subsection{Energetic Variational Approach}

We start with a brief introduction to the classical energetic variational approach (EnVarA), which was developed from the variational principle, proposed by Rayleigh \cite{strutt1871some} for purely frictional systems, that Onsager tried to extend to physical systems in general \cite{onsager1931reciprocal, onsager1931reciprocal2}.  

The starting point of an energetic variational approach is a prescribed energy-dissipation law for an isothermal and closed system, which comes from the first and second law of thermodynamics \cite{giga2017variational}. Indeed, for a thermodynamic process without transfer of matter, the first law of thermodynamics is often formulated as
  \begin{equation}
\frac{\dd}{\dd} (\mathcal{K} + \mathcal{U}) = \dot{W} + \dot{Q}, 
  \end{equation}
  that is the rate of change of the kinetic energy $\mathcal{K}$ and the internal energy $\mathcal{U}$  can be attributed to either the work $\dot{W}$ done by  the external environment or the heat $\dot{Q}$. To analyze heat, one needs to introduce the entropy $\mathcal{S}$, which satisfies a time dependent version of the second law of thermodynamics:
  \begin{equation}
    T \frac{\dd \mathcal{S}}{\dd t} = \dot{Q} + \triangle, \quad \triangle \geq 0,
  \end{equation}
  where $T$ is the temperature and $\triangle$ is the entropy production. Subtracting the two laws, one arrives at an energy-dissipation law
\begin{equation}\label{ED_01}
\frac{\dd}{\dd t} E^{\text{total}}(t) = - \mathcal{D}(t),
\end{equation}
for isothermal and closed system ($\dot{W} = 0$).
Here $E^{\text{total}}$ is the total energy, which is the sum of the Helmholtz free energy $\mathcal{F} = \mathcal{K} - T \mathcal{S}$ and the kinetic energy $\mathcal{K}$. $\mathcal{D}$ is the rate of energy dissipation that is related to the entropy production. 

For a given energy-dissipation law, the energetic variational approach provides a paradigm to determine the dynamics of system through two distinct variational processes: the Least Action Principle (LAP) and the Maximum Dissipation Principle (MDP) \cite{liu2009introduction,giga2017variational}. Specifically, the Least Action Principle states that the dynamics of a Hamiltonian system is determined by a critical point of the action functional $\mathcal{A}(\x) = \int_0^T \mathcal{K} - \mathcal{F} \dd t$ with respect to $\x$ (the trajectory in Lagrangian coordinates, if applicable) \cite{arnol2013mathematical, giga2017variational}, i.e., 
\begin{equation}
  \delta \mathcal{A} =  \int_{0}^T \int_{\Omega} (f_{\text{inertial}} - f_{\text{conv}})\cdot \delta \x  \dd \x \dd t,
\end{equation}
where $f_{\rm inertial}$ is the inertial force and $f_{\text{conv}}$ is the conservative force.
Formally, the LAP represents the fact that force multiplies by distance is equal to the
work, i.e., $\delta E = \text{force} \times \delta x$,
where $x$ is the location, $\delta$ represents the variation/differential.
In the meantime, for a dissipative system ($\mathcal{D} \geq 0$), we follow Onsager \cite{onsager1931reciprocal, onsager1931reciprocal2} and determine the dissipative force $f_{\rm diss}$ by minimizing the dissipation functional $\mathcal{D}$ with respect to the ``rate'' $\x_t$, known as the Maximum Dissipation Principle (MDP), i.e.,
\begin{equation}
\delta \left(\frac{1}{2}\mathcal{D} \right) = \int_{\Omega} f_{\text{diss}} \cdot \delta \x_t~ \dd \x.
\end{equation}
The dissipation $\mathcal{D}$ is often assumed to be quadratic in terms  of the ``rate'' $\x_t$ \cite{doi2011onsager}, that is 
\begin{equation}\label{classic_D}
\mathcal{D}[\bm{x}, \bm{x}_t] =  \int \mathcal{G}(\bm{x}) \bm{x}_t \cdot \bm{x}_t  \dd \x,
\end{equation}
where $\mathcal{G} (\bm{x})$ is a positive semi-definite matrix for given $\bm{x}$. The assumption (\ref{classic_D}) corresponds to the linear response theory in non-equilibrium thermodynamics \cite{onsager1931reciprocal, onsager1931reciprocal2, de2013non}.

According to force balance (Newton's second law, in which the inertial force plays the role of $ma$), we have
\begin{equation}\label{FB}
\frac{\delta A}{\delta \x} = \frac{1}{2}\frac{\delta \mathcal{D}}{\delta \x_t},
\end{equation}
which defines the dynamics of the system.

The EnVarA framework shifts the main task of modeling to the construction of the energy-dissipation law.
As an illustration, we consider a simple example originally proposed by Lord Rayleigh \cite{strutt1871some}, a spring-mass system, in which a Hookean spring of which one end is attached to a wall and another end to a mass $m$. Then,
\begin{equation*}
  \mathcal{K} = \frac{m}{2} x_t^2, \quad \mathcal{F} = \frac{k}{2} x^2, \quad \mathcal{D} = \gamma  x_t^2, 
\end{equation*}
where $k$ is the spring constant, and $\gamma$ is damping coefficient. The corresponding action functional is defined as
\begin{equation*}
\mathcal{A} = \int_{0}^T \frac{m}{2} x_t^2 - \frac{k}{2} x^2 \dd t.
\end{equation*}
Then the LAP, i.e. variation of $\mathcal{A}$ with respect to the trajectory $x(t)$ gives rise to
\begin{equation}
\frac{\delta \mathcal{A}}{\delta x} = - m x_{tt} - k x.
\end{equation}
Meanwhile, the MDP, taking the variation of $\mathcal{D}$ with respect to $x_t$ gives
\begin{equation}
\frac{1}{2}\frac{\delta \mathcal{D}}{\delta x_t} = \gamma x_t.
\end{equation}
Hence, the force balance condition (\ref{FB}) yields
\begin{equation}
m x_{tt} + k x + \gamma x_t = 0.
\end{equation}
In an overdamped case ($m \ll \gamma$), the $m x_{tt}$ term can be neglected \cite{schuss1980singular}, and the system becomes a gradient flow with the dynamics given by
\begin{equation*}
x_t = - \frac{1}{\gamma} \frac{\delta \mathcal{F}}{\delta x},
\end{equation*}
In the following, we always working on the overdamped region, and neglect the kinetic energy in (\ref{ED_01}).

\subsection{EnVarA with chemical reaction}

As well as any other variational principles, classical energetic variational approaches deal with mechanical systems, which are indeed based on the Newton's second law $F = ma$. In general, chemical reactions cannot be understood from Newtonian mechanics, as there is no clear mechanical interpretation for the chemical potential \cite{ge2016mesoscopic}.

Many papers try to build an Onsager type variational theory for chemical reaction systems \cite{wei1962axiomatic, van1962general, feinberg1972complex, bataille1978nonequilibrium, mielke2011gradient, mielke2016generalization, beris1994thermodynamics, kondepudi2014modern}.%
For example, \citeauthor{mielke2011gradient}  
established the gradient flow structure for reaction-diffusion systems with reversible mass-action kinetics by using the dual dissipation potential \cite{mielke2011gradient}.
 As an extension of the  GENERIC framework, \citeauthor{grmela2012fluctuations} showed the geometry associated with the law of mass action is the the contact geometry. He extended the mass-action kinetics to  account  for  the  influence  of  inertia  and  fluctuations, which can be adopted to complicated reaction systems involving many intermediate reactions \cite{grmela2012fluctuations}.

For the reaction (\ref{rev_reaction}) with the law of mass action, it has been discovered for a long time that there exists a Lyapunov functional \cite{shear1967analog,  Desvillettes2006, mielke2011gradient, grmela2012fluctuations, anderson2015lyapunov, perthame2015parabolic, ge2016mesoscopic, mielke2017non}, which is the free energy of the system.
 The free energy can be written down in various equivalent form; here we adopt a thermodynamics based form 
 \begin{equation}\label{Energy_U_ABC}
  \begin{aligned}
    & \mathcal{F}(c_A, c_B, c_C ; U_A, U_B, U_C) \\
    & = \int_{\Omega} \RT \Bigl( c_A (\ln c_A  - 1 ) + c_B (\ln c_B - 1) + c_C (\ln c_C - 1) \Bigr) \\
    & \qquad ~ + c_A U_A + c_B U_B + c_C U_C ~ \dd \x, \\
  \end{aligned}
\end{equation}
for the chemical reaction (\ref{rev_reaction}). The first three terms in (\ref{Energy_U_ABC}) form the free energy of a mixture of ideal gases without chemical reactions, which corresponds to the entropy. Indeed, for a mixture of ideal gases with $N$ species, the chemical potential of a substance $j$ is expressed by \cite{lebon2008understanding}
\begin{equation}\label{c_1}
  \mu_j = \mu^{0} + \RT \ln x_j,
\end{equation}
where $\mu^{0}$ is the reference chemical potential, and $x_j$ is the concentration of the substance $j$.
Since the chemical potential is defined relative to its value at an arbitrary reference state, we can take $\mu^0 = 0$. The free energy of the mixture of ideal gases, corresponding to the chemical potential (\ref{c_1}) with $\mu_0 = 0$, is given by
\begin{equation}
  \mathcal{F}[x_i] = \int_{\Omega} \RT \sum_{i = 1}^N x_i (\ln x_i - 1) \dd x.
\end{equation}
The last three terms in  (\ref{Energy_U_ABC}) can be viewed as internal energies  stored inside the molecular $A$, $B$ and $C$. In the case without chemical reaction, since $c_A$, $c_B$ and $c_C$ do not change with respect to time, these terms are constants that can be ignored. 
From a modeling perspective, as also pointed out in \cite{grmela2012fluctuations}, $U_i$ are parameters that determine the equilibrium of the system. For the given free energy $\mathcal{F}(c_A, c_C, c_C ; U_A, U_B, U_C)$ defined in (\ref{Energy_U_ABC}), the corresponding chemical potential of each species is given by
\begin{equation}
\mu_i = \frac{\delta \mathcal{F}}{\delta c_i} = \RT \ln c_i + U_i, \quad i = A, B, C
\end{equation}
At a chemical equilibrium, the chemical potential of both sides of the reaction are equal, i.e., the {\emph affinity} 
\begin{equation}
  \gamma \mu_C - \alpha \mu_A - \beta \mu_c = 0,
\end{equation}
which indicates that
\begin{equation}
\ln \frac{(c_A^{\infty})^{\alpha} (c_B^{\infty})^{\beta}}{(c_C^{\infty})^{\gamma}} =  \frac{1}{\RT} (\gamma U_C - \alpha U_A - \beta U_B) := \frac{\Delta U}{\RT}.
\end{equation}
Here $\Delta U = \gamma U_C - \alpha U_A - \beta U_B$ is the difference of internal energy between the state $\{\alpha A, \beta B\}$ and the state $\{ \gamma C \}$. Then the equilibrium constant $K_{eq}$ is defined as \cite{keener1998mathematical} 
\begin{equation} \label{Def_Keq}
  K_{eq} \triangleq \frac{(c_A^{\infty})^{\alpha} (c_B^{\infty})^{\beta}}{(c_C^{\infty})^{\gamma}}= e^{\frac{\Delta U}{\RT}} ,
\end{equation}
which is an exponential representation of the difference in internal (`chemical') energies. 

In our approach, we always assume the existence of the free energy $\mathcal{F}$, which is different from most of previous approaches. Those approaches start with the mass-action kinetics and show the existence of the free energy under the \emph{detailed balance} condition \cite{Desvillettes2006, mielke2011gradient}.
For a general system, the free energy $\mathcal{F}$ might contains various different mechanism and cannot be derived by mathematics alone until a physical model is specified. Here we assume that $U_i$ are constants to illustrate our approach. Confrontation with real experimental data will undoubtedly motivate more complex models. It should be emphasized that the choice of the free energy $\mathcal{F}$ determines the chemical equilibrium (if it exists) of the system.

As pointed out in \cite{oster1974chemical}, one of the difficulties in applying variational principles to a chemical reaction arose from the stoichiometric constraint (\ref{constraint}).
To overcome this difficulty, \citeauthor{oster1974chemical} treated the reaction kinetics in a differential geometric context and introduced the ``reaction trajectory''. 
The idea of using reaction trajectory, also known as \emph{extent of reaction} or \emph{degree of advancement}, as a new stable variable can be traced back to De Donder \cite{kondepudi2014modern},
 and has been used for both deterministic and stochastic descriptions of chemical reactions for a long time \cite{reti1984analogies, keizer2012statistical, anderson2015lyapunov}.
Roughly speaking, a reaction trajectory accounts for the ``number'' of forward chemical reactions that has occurred by time $t$. 
By introducing the reaction trajectory $R(t)$, the concentrations of $A, B$ and $C$ for the single chemical reaction (\ref{rev_reaction}) are given by
\begin{equation}\label{Kinematic_ABC}
  c_i(t) = c_i(0) + \sigma_i R(t),
\end{equation}
which can be viewed as the kinematics of the chemical reaction that embodies the constraint (\ref{constraint}).

By using the reaction trajectory, we can reformulate the free energy  $\mathcal{F}(c_A, c_B, c_C ; U_A, U_B, U_C)$ defined in (\ref{Energy_U_ABC}) in terms of $R(t)$, and use the energy-dissipation law
\begin{equation}\label{ED_R}
\frac{\dd}{\dd t} \mathcal{F}[R ; U_A, U_B, U_C] = - \mathcal{D}[R, R_t],
\end{equation}
to model the reaction kinetics of the chemical reaction (\ref{rev_reaction}).
Here $\mathcal{D}[R, R_t]$ is a dissipation of the system. Different choices of $D[R, R_t]$ determine different reaction kinetics. 

Unlike mechanical systems, chemical reactions are often far from thermodynamic equilibrium, so the dissipation $D[R, R_t]$ may not be quadratic in terms of $R_t$ \cite{de2013non, beris1994thermodynamics}. In order to deal with the general form of the dissipation, we need to extend the classical EnVarA. Assume $\mathcal{D}(R, R_t)$ takes the form 
\begin{equation}
\mathcal{D}[R, R_t] = \left( \Gamma(R, R_t), R_t  \right) \geq 0,
\end{equation}
where $(.,.)$ is an inner product,
since
\begin{equation}
  \frac{\dd}{\dd t} \mathcal{F}[R]  = \left(\frac{\delta \mathcal{F}}{\delta R}, R_t \right),
\end{equation}
the energy-dissipation law (\ref{ED_R}) implies
\begin{equation}\label{eq_GR}
  \Gamma(R, R_t) =  - \frac{\delta \mathcal{F}}{\delta R},
\end{equation}
which is the equation for the chemical kinetics.
Interestingly, notice that
\begin{equation}
  \frac{\delta \mathcal{F}}{\delta R} = \sum_{i=1} \sigma_i \mu_i,
\end{equation}
is exactly the \emph{affinity} of chemical reaction, as defined by De Donder \cite{de1927affinite, de1936thermodynamic}. The affinity plays a role of the ``force'' that drives chemical reactions, and $R_t$ can be identified as the reaction velocity (or \emph{rate of conversion} \cite{kondepudi2014modern}). Just as in a mechanical system, the dissipation of this chemical reaction system gives the relation between the reaction velocity $R_t$ and the chemical force. Next we discuss two typical choices of the dissipations.

\noindent{\bf General Law of mass action}: The law of mass action can be derived from the energy-dissipation law (\ref{ED_R}) by choosing 
\begin{equation}\label{Dissipation_1}
  \mathcal{D}[R, R_t]  = \RT~ R_t \ln \left( \frac{R_t}{k_r c_C^{\gamma}} + 1 \right).
\end{equation}
Indeed, the energetic variational procedure gives
\begin{equation}
\RT \ln \left( \frac{R_t}{k_r c_C^{\gamma}} + 1 \right) = - \frac{\delta}{\delta R} \mathcal{F} [R].
\end{equation}
Notice that
\begin{equation*}
  \begin{aligned}
    \frac{\delta}{\delta R} \mathcal{F} [R]  
     = \RT \ln \left( \frac{c_c^{\gamma}}{c_A^{\alpha} c_B^{\beta}} \right) - \alpha U_A - \beta U_B + \gamma U_C, \\
  \end{aligned}
\end{equation*}
which indicates that 
\begin{equation}\label{eq_R}
 \ln \left( \frac{R_t}{k_r c_C^{\gamma}} + 1 \right) = \ln \left( \frac{c_A^{\alpha} c_B^{\beta}}{c_C^{\gamma}} \right) - \frac{\Delta U}{\RT},    
\end{equation}
where the right-hand side is determined by the difference of internal energy $\Delta U$ between the state $\{\alpha A, \beta B\}$ and the state $\{ \gamma C \}$. Although (\ref{eq_R}) looks complicated, direct computation shows that
\begin{equation}\label{rate_1}
  \begin{aligned}
    R_t & = k_r c_C^{\gamma} \left( \frac{1}{K_{eq}} \frac{ c_A^{\alpha} c_B^{\beta} }{c_C^{\gamma}} - 1 \right) = k_f c_A^{\alpha} c_B^{\beta} - k_r c_C^{\gamma}, \\
  \end{aligned}
\end{equation}
which is the classical law of mass action. Here the relation $K_{eq} = e^{\frac{\Delta U}{\RT}}  = \frac{k_r}{k_f}$ is used to get the last equality. 
It is worth mentioning that the dissipation (\ref{Dissipation_1}) is identical to a widely used form of the entropy production \cite{ge2016mesoscopic, ge2017mathematical}
    \begin{equation*}
   \triangle = (r_{f} - r_{r})\ln \left( \frac{r_{f}}{r_{r}} \right),
    \end{equation*}
  where $r_{f}$ and $r_{r}$ are forward the reverse reaction rates. 

As a generalization of (\ref{Dissipation_1}), we can consider a more general form of the dissipation
  \begin{equation}\label{Dissipation_g}
    \mathcal{D}[R, R_t] = \eta_1(R)  R_t \ln( \frac{R_t}{\eta_2 (R)} + 1),
  \end{equation}
  where $\eta_1(R) > 0$ and $\eta_2 (R) > 0$, then $\mathcal{D} [R, R_t] \geq 0$ for the admissible $R$. By choosing $\eta_1(R)$ and $\eta_2(R)$ properly, we can have a concentration dependent reaction rate, which is often used to provide a thermodynamic description of an autocatalytic chemical reaction \cite{keener1998mathematical}.

  \noindent {\bf Linear Response Theory:}
  In nonequilibrium thermodynamics, it is often assumed that the dissipation of the total energy is
  a quadratic function of ``rate'' of change of state variables, which is known as the linear response theory. \cite{onsager1931reciprocal, onsager1931reciprocal2, de2013non}.
In our case, the linear response theory gives a form of the dissipation term
  \begin{equation}\label{Dissipation_2}
    \mathcal{D} [R, R_t] = \eta(R) |R_t|^2.
  \end{equation}
  Then the variational procedure gives
  \begin{equation*}
    \begin{aligned}
      \eta(R) R_t & = - \frac{\pp }{\pp R} \mathcal{F} [R] =  \RT \ln \left( \frac{c_C^{\gamma}}{c_A^{\alpha}c_B^{\beta}} \right) - \Delta U. 
    \end{aligned}
  \end{equation*}
  By choosing $\eta(R) =  \RT$, the reaction rate is given by
  \begin{equation}\label{rate_2}
    \begin{aligned}
      r  = R_t  
      &  = \ln \left( \frac{1}{K_{eq}} \frac{c_A^{\alpha}c_B^{\beta}}{c_C^{\gamma}}  \right), \\ 
    \end{aligned}
  \end{equation}
  a form of which is more complicated than the law of mass action. 
  \begin{remark}
    The law of mass action gives a simple form of the reaction rate $r$ in terms of concentrations, however, the dissipation in terms of $R$ and $R_t$ becomes complicated (See eq. (\ref{Dissipation_1})). On the other hand, if the dissipation is taken to be simple that described by linear response theory, the the reaction rate $r$ becomes complicated (See eq. (\ref{rate_2}) ). 
    \end{remark}

Some early variational treatments of chemical reactions are based on the linear response assumption \cite{biot1977variational}, which arose from the near equilibrium assumption. Indeed, for the chemical reaction, we have $R_t \approx 0$  near the equilibrium, then the Taylor expansion gives us
  \begin{equation}\label{Q_log_app}
    \eta_1(R) \ln \left( \frac{R_t}{\eta_2 (R)} + 1 \right) \approx \frac{\eta_1(R)}{\eta_2(R)} |R_t|^2.
  \end{equation}
  Thus, one can view the dissipation (\ref{Dissipation_2}) as a linear approximation near equilibrium to (\ref{Dissipation_g}).
  However, it is believed that, except for the special case that is close to equilibrium, the driving force for chemical reaction is a nonlinear functional of the system variables \cite{de2013non, beris1994thermodynamics}.

It is straightforward to extend the above EnVarA description to a general \emph{reversible} chemical reaction system contains $N$ species $\{ X_1, X_2, \ldots X_N \}$ and $M$ reactions,
given by
\begin{equation*}
\ce{ $\alpha_{1}^{l} X_1 + \alpha_{2}^{l}X_2 + \ldots \alpha_{N}^{l} X_N$ <=> $\beta_{1}^{l} X_1 + \beta_{2}^{l}X_2 + \ldots \beta_{N}^{l} X_N$}, 
\end{equation*}
for $l = 1, \ldots, M$. Let  $\bm{c} = (c_1, c_2, \ldots, c_N)^{\rm T} \in \mathbb{R}^N$ be the concentrations of all species. The kinematics of the system are then given by
\begin{equation}
{\bm c} = {\bm c}_0 + {\bm \sigma} {\bm R},
\end{equation}
where ${\bm c}_0$ is the initial concentrations, ${\bm R} \in \mathbb{R}^M$ represents $M$ reaction trajectories of $M$ reactions, $\bm{\sigma} \in \mathbb{R}^{N \times M}$ with $\sigma_{il} = \beta^l_i - \alpha^l_i$ is the stoichiometric matrix. The reaction kinetics of this chemical reaction network can be described by the energy-dissipation 
\begin{equation}
\frac{\dd}{\dd t} \mathcal{F}[{\bm c}({\bm R})] = - \mathcal{D}[{\bm R}, \pp_t {\bm R}],
\end{equation}
where 
\begin{equation}
  \mathcal{F}[{\bm c}] = \sum_{i=1}^N c_i (\ln c_i - 1) + c_i U_i, 
\end{equation}
with $U_i$ be the internal energy, and the dissipation can be taken as
\begin{equation}
  \mathcal{D}[{\bm R}, \pp_t {\bm R}] = - \sum_{l = 1}^M \pp_t R_l \ln \left( \frac{\pp_t R_l}{\eta_l({\bm c}({\bm R}))} + 1 \right)
\end{equation}
to be consistent with mass action kinetics. Then the variational procedure gives the dynamics of the chemical reaction
\begin{equation}
  \ln \left( \frac{\pp_t R_l}{\eta_l({\bm c}({\bm R}))} + 1 \right) = - \frac{\delta F}{\delta R_l},
\end{equation}
where $\frac{\delta F}{\delta R_l}$ is the affinity of the $l$-th chemical reaction. 




\subsection{Reaction-Diffusion System}
The above EnVarA description of a chemical reaction provides a way to couple chemical reactions with other mechanical mechanisms, such as diffusion and electro-diffusion, in a unified variational framework. As an illustration, we apply the EnVarA to a reaction-diffusion system, which is a simple example of a mechano-chemical or chemo-mechanical system. Reaction-diffusion type partial differential equations are used widely to model biological processes \cite{perthame2015parabolic}, such as molecular motors \cite{julicher1997modeling}, prion diseases \cite{fornari2019prion}, and tumor growth \cite{hawkins2012numerical}.

Consider a reaction-diffusion system in a fixed domain $\Omega$ with the reaction given by (\ref{rev_reaction}), then the kinematics for the concentrations $c_A$, $c_B$ and $c_C$  are given by
\begin{equation}\label{Kinematic_RD}
\pp_t c_i(\x, t) + \nabla \cdot (c_i \uvec_i)  = \sigma_i \pp_t R(\x, t), \quad i = A, B, C
\end{equation}
where $\uvec_i$ is the macroscopic velocity of different species induced by the diffusion process, $R$ is the reaction trajectory for the chemical reaction (\ref{rev_reaction}).

The energy-dissipation law of the reaction-diffusion system can be formulated as
\begin{equation}\label{ED_RD}
  \begin{aligned}
    & \frac{\dd}{\dd t} \mathcal{F}(c_A, c_B, c_C) =  - (\mathcal{D}_{\rm chem} +  \mathcal{D}_{\rm mech}) \\
  \end{aligned}
\end{equation}
where the free energy $\mathcal{F}(c_A, c_B, c_C)$ is given by Eq. \eqref{Energy_U_ABC}, which is same as for a pure reaction system. $\mathcal{D}_{\rm chem}$ is the dissipation arises from the chemical reaction, which is given by $\mathcal{D}_{\rm chem} = (\Gamma(R, R_t), R_t) \geq 0$ as in the last subsection. $\mathcal{D}_{\rm mech}$ is the dissipation due to the diffusion process, which is often taken as \cite{liu2019lagrangian}
\begin{equation*}
  \mathcal{D}_{\rm mech} = \int \eta_A({\bm c}) |\uvec_A|^2 +  \eta_B({\bm c}) |\uvec_B|^2 +  \eta_C({\bm c}) |\uvec_C|^2 \dd \x
\end{equation*}
to model the friction of the fluid fluxes. It is important to notice that in this case, the dynamics of both the mechanical and chemical parts are derived from the same free energy.

Notice that 
\begin{equation}
  \frac{\dd}{\dd t} \mathcal{F}(c_A, c_B, c_C) = \sum_{i=1}^3 (\nabla \mu_i, \uvec) +  \left(\frac{\delta \mathcal{F}}{\delta R}, R_t \right), 
\end{equation}
by using the generalized energetic variational approach, the equations for $R$ and $\uvec_i$ can be derived as
\begin{equation}\label{eq_JR}
 \begin{cases}
   & \Gamma(R, R_t) =  - \dfrac{\delta \mathcal{F}}{\delta R}  \\
   & \\
    & \eta_i({\bm c}) \uvec_i = - c_i \nabla \left(\dfrac{\delta \mathcal{F}}{\delta \bm{c}_i} \right), \quad i = A, B, C.  \\
  \end{cases}
\end{equation}
Here the first equation is the same as (\ref{eq_GR}), and the second equation is actually Fick's Law of diffusion \cite{giga2017variational}.
By choosing $\eta_i({\bm c}) = c_i (i = A, B, C)$ and combining (\ref{eq_JR}) with (\ref{Kinematic_RD}), we can obtain a reaction-diffusion system
\begin{equation*}
    \pp_t c_i = \nabla \cdot (\nabla c_i) + \sigma_i r(\x, t) \\
\end{equation*}
where $r(\x, t)$ is the reaction rate determined by the choice of $D_{\rm chem}(R, \pp_t R)$ as discussed in the last subsection.

\begin{remark}
  It is worth mentioning that here we only couple the chemical reaction with dissipative
  mechanics (e.g. diffusion). The chemical and mechanical parts share the same free energy but have different dissipation mechanisms \cite{biot1982thermodynamic, liero2013gradient}. 
  In \cite{klika2013coupling}, the authors develop a novel approach that couples chemical kinetics with non-dissipative time reversible mechanics, such as elastic deformations, which has potential applications in biology. We refer interested readers to \cite{klika2013coupling, pavelka2018multiscale} for the mathematical formula of such a type of coupling.
\end{remark}

\section{Input-Output Relation and the Dissipation}\label{sec:out}

\begin{figure}[!hbt]
  \centering
  \includegraphics[width = \linewidth]{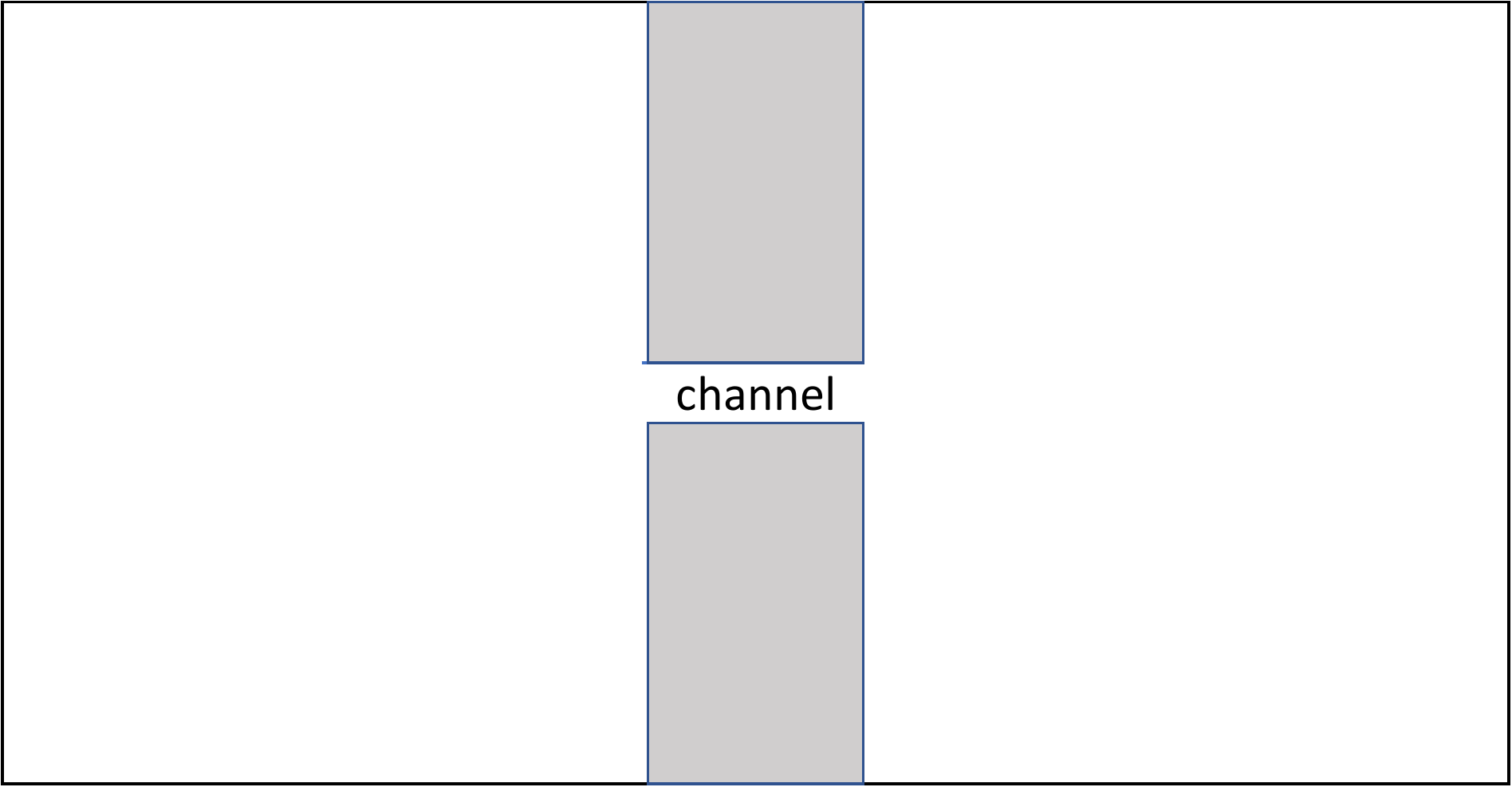}
  \caption{Setup of an open nonequilibrium system}\label{setup}
\end{figure}

As pointed out previously, in the EnVarA framework, the dynamics of chemical reactions are determined by the choice of the dissipation. Notice that the equilibrium constant $K_{eq}$ 
is determined by the choice of the free energy,
measurements of just $K_{eq}$ cannot distinguish different dissipation mechanisms. 
In the meantime, although chemical reactions are believed to operate far away from equilibrium \cite{de2013non, beris1994thermodynamics}, directly simulating the ODE system for the two dissipations (\ref{rate_1}) and  (\ref{rate_2}) produce almost identical results since both systems move to the equilibrium so quickly. To distinguish different reaction kinetics, it is necessary to study a non-equilibrium system, which can predict different dependence of rate on concentrations and different time courses of the chemical reaction. 
In this section, we study a particular setup, shown in Fig. \ref{setup}, that can be realized in experiments.

Such a setup is chosen to give reproducible input-output functions for different dissipations. In this system, a narrow channel connects two bath, as shown in Fig. \ref{setup}.
We assume the chemical reaction 
\begin{equation}
\ce{A + B <=> C} 
\end{equation}
happens inside the channel, and the average concentrations of $A$ and $B$ in the left bath can be maintained by the boundary condition. The species in the left bath are sources, and the species in the right bath are outputs.
The chemical reaction is the ``transfer function''. The sources provided by the ``left bath'' can keep the system away from the equilibrium.

\begin{figure}[!htb]
  \begin{center}
     \begin{overpic}[width = \linewidth]{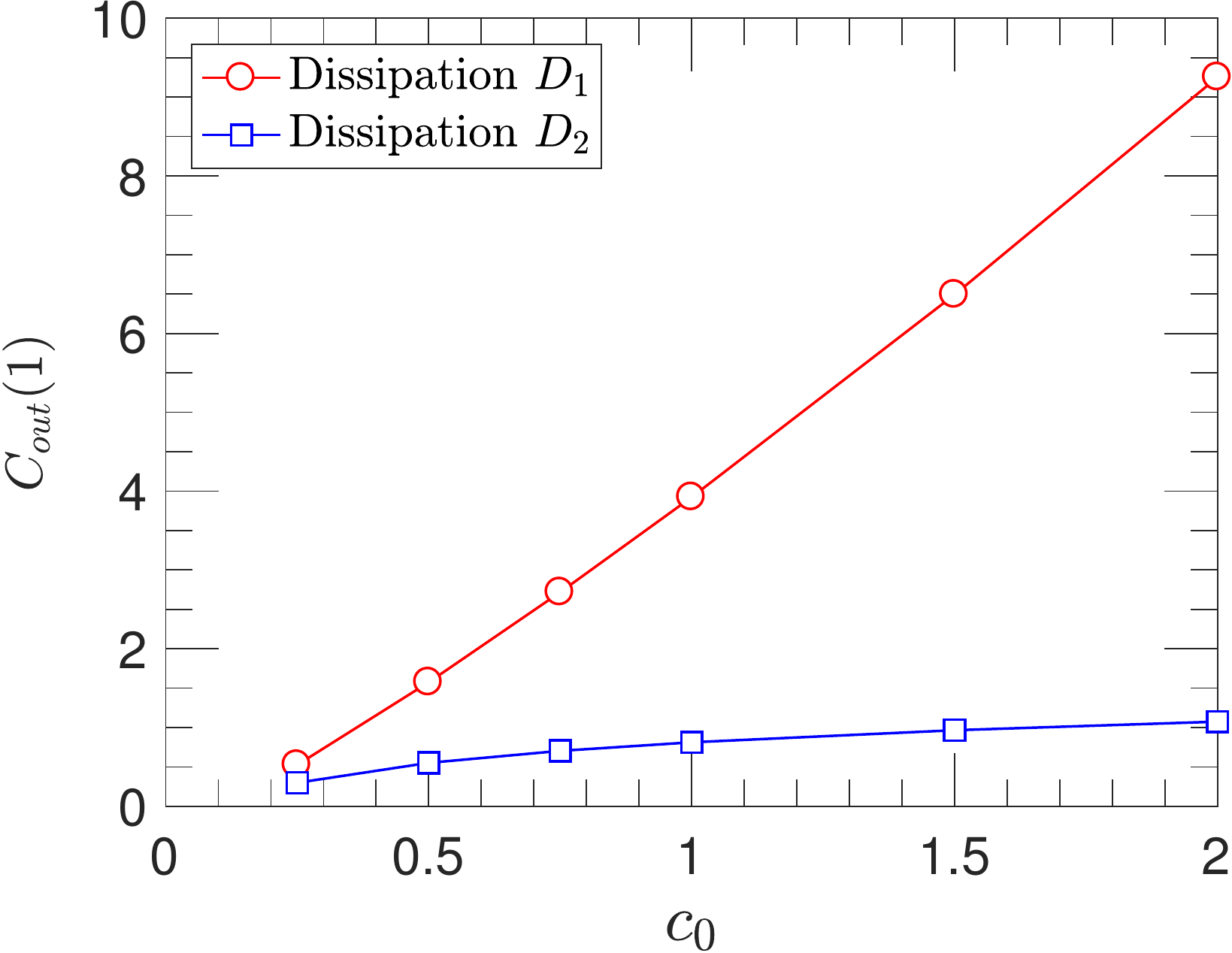}
   \end{overpic}
   \end{center}
   \caption{The output $C_{out}(t)$ as a function of input $c_0$ when $t = 1$ for dissipation (\ref{D_1}) [circle] and dissipation (\ref{D_2}) [square].}\label{In_Out}
 \end{figure}

 \begin{figure*}[!hbt]
  \centering
  \includegraphics[width = \linewidth]{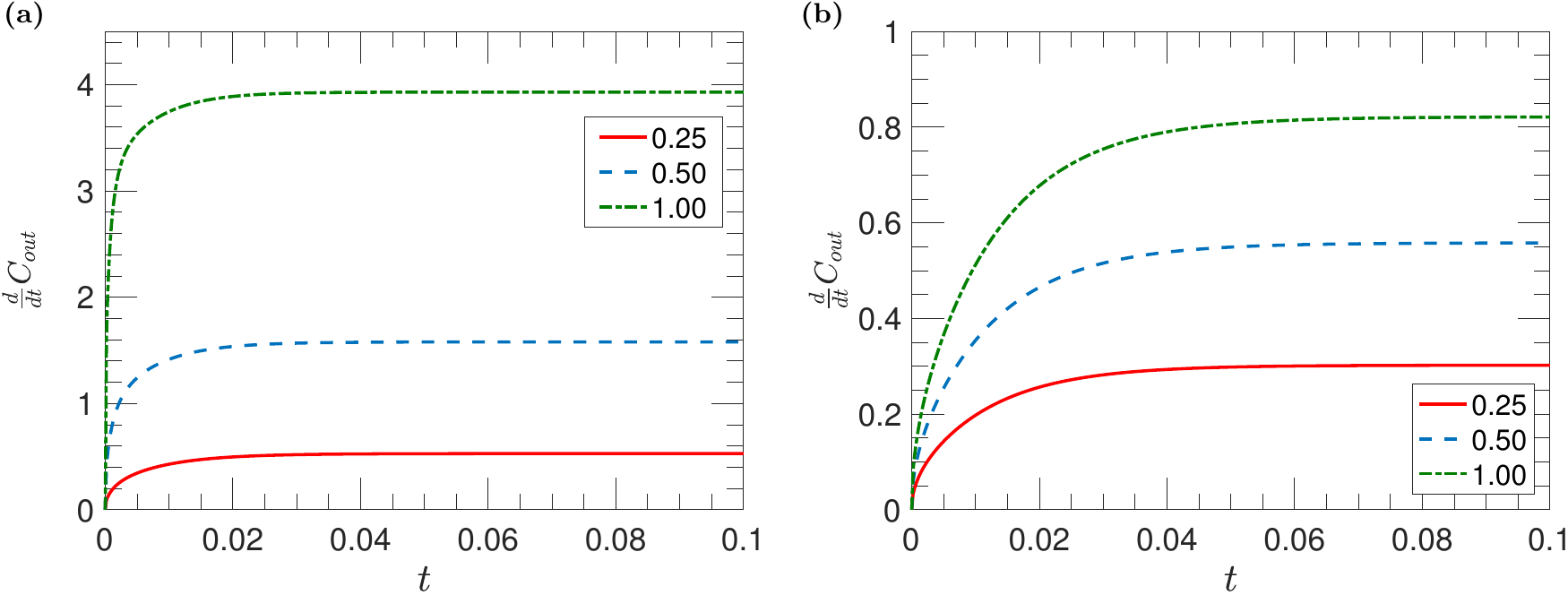}
  \caption{ $\frac{\dd}{\dd t} C_{out}$ as a function of $t$ for two dissipations for various of $c_0$ ($c_0 = 1, 0.5$ and $0.25$ from top to bottom in each figure). (a) Dissipation (\ref{D_1}), (b) Dissipation (\ref{D_2}). }\label{Res_D12}
\end{figure*}

This system can be viewed as an abstract representation of one component of complex biological networks, in which a enzyme localizes a particular chemical reaction and moves the reactants into products. This representation links chemical reactions to the two terminal devices of electrical and electronic engineering \cite{perelson1974chemical, alon2003biological, feinberg2019foundations,rao2016nonequilibrium}. Each reaction is a separately defined device (loosely speaking) with an input and output and its own input-output relations. The enzymes can be thought as two terminal devices, as diodes, that move reactants into products, from one chemical state to another, much as channels are diodes that move ions from one physical location to another through a reaction path \cite{perelson1974chemical, eisenberg1990channels}. Although treating chemical reaction systems by electric circuit theory has existed for a long time, the spatial effect seems to be overlooked. Reactions in biology occur in different physical locations, so the products of one enzyme's reaction move, by diffusion (and perhaps migration and convection), to become reactants for the reaction catalyzed by an enzyme in a different location.

Mathematically, since the channel is very narrow, we can treat this problem as one one-dimensional, with the domain given by $[-\epsilon, \epsilon]$. We fix $\epsilon = 0.1$ through this section.
As mentioned previously, the concentrations of $A$ and $B$ in the left bath are maintained, which gives us the Dirichlet boundary conditions of $A$ and $B$ in the left-end of the channel.
 We can impose the boundary conditions
\begin{equation*}
  \begin{aligned}
    & c_A(- \epsilon, t) = c_0, \quad \pp_{x} c_A(\epsilon, t) = 0, \\
    & c_B(- \epsilon, t) = c_0, \quad \pp_{x} c_B(\epsilon, t) = 0, \\
    & \pp_{x} c_C(- \epsilon, t) = 0, \quad  c_C(\epsilon, t) = 0.1, \\
  \end{aligned}
\end{equation*}
and treat $c_0$ as the single input of our system.

Since $c_C$ satisfies the Dirichlet boundary condition on the right-end of the channel, we can define the amount of $C$ diffuse into right bath by time $T$ as
    \begin{equation}\label{Amount_C}
      C_{out}(T) = \int_{-\epsilon}^{\epsilon} R(x, T) \dd x - \int_{-\epsilon}^{\epsilon} (c(x, T) - c_0(x)) \dd x,
    \end{equation}
  which is the output of our system. The flux of $C$ or the rate of change of amount of $C$ in the right bath is defined as $\dfrac{\dd}{\dd t} C_{out}$. The initial concentrations of $A$, $B$ and $C$ in the channel are constants $c_A^0(\x) = c_B^0(\x) = c_0$ and $c_C^0(\x) = 0.1$.

We fix $K_{eq} = 0.1$ and assume 
the free energy is given by
\begin{equation*}
  \begin{aligned}
  \mathcal{F} = \int & c_A \ln \left(0.1 c_A\right) - 1) \\
  & + c_B (\ln \left( 0.1 c_B \right) - 1) + c_C \ln \left( c_C \right) - 1) \dd \x. \\
  \end{aligned}
\end{equation*}
We focus on two types of dissipations, a generalized law of mass action
\begin{equation}\label{D_1}
  D_1(R, \pp_t R) =  R_t \ln ( R_t + 1),
\end{equation}
and a dissipation based on the linear response assumption
\begin{equation}\label{D_2}
D_2(R, \pp_t R) = |R_t|^2.
\end{equation}
These two dissipations (\ref{D_1}) and (\ref{D_2}) are almost same near a equilibrium (see (\ref{Q_log_app})). By numerical simulations, we show that the input-output relation depends on the choice of the dissipation in this nonequilibrium setup.

  


Fig. \ref{In_Out} shows the output $C_{out}(t)$ as a function of the input $c_0$ at $t = 1$ for the two choices of dissipation. For small $c_0$, the outputs are nearly same for the two dissipation functionals. However, the output for dissipation (\ref{D_1}) is much larger than that for the dissipation (\ref{D_2}) when $c_0$ is large. Formally, from the computations in Sec. \ref{set:EnVarA}, we know $R_t = \ln \left( \frac{1}{K_{eq}} \frac{c_A c_B}{c_C}\right)$, for the dissipation (\ref{D_1}), while $R_t = \frac{1}{K_{eq}} \frac{c_Ac_B}{c_C} - 1$ for the dissipation (\ref{D_2}). For $c_0 = 0.1$, the system is at the equilibrium, so $C_{out} = 0$. When $c_0$ is large, the dissipation (\ref{D_2}) will determine a larger reaction rate.

We also consider $\frac{\dd}{\dd t} C_{out}$ as a function of $t$ for two choices of dissipation for various of $c_0$. The results are shown in Fig. \ref{Res_D12}.
The time courses in Fig. \ref{Res_D12} show that for different dissipations and different inputs, $\frac{\dd}{\dd t} C_{out}$ tends to a constant, which is a function of input for a given dissipation. 

Although the dissipation (\ref{D_1}) and (\ref{D_2}) are almost the same when near equilibrium, the above simulations indicate that in a non-equilibrium setting, the input-output relationship might be very different for different choices of dissipations since the system is maintained far from equilibrium due to inputs of reactants through the boundary condition.
This suggests that one might be able to determine the dissipation through experimental measurements and solving the inverse problem \cite{burger2007inverse}.


\section{Summary}
In this paper, we apply a generalized energetic variational approach (EnVarA) to a reversible chemical reaction system, which enables us to couple chemical reactions with other mechanical effects, such as diffusion, as well as the thermal effect. In our approach, the chemical equilibrium (if it exists) is determined by the choice of the free energy, and the dynamics of a chemical reaction is determined by the choice of the dissipation. The classical law of mass action can be derived through a particular form of the dissipation. 

To distinguish different dissipations, we study a non-equilibrium reaction-diffusion system with boundary effects.
This system can be viewed as an abstract representation of a building block of complex biological networks, in which a enzyme that localizes a particular  chemical  reaction and   moves  reactants  into  products. Our simulation results show that the input-output of such a system depends on the choice of the dissipation. If the experimental system is reasonably reproducible, the dissipation mechanism can be obtained by experimental measurements and studying an inverse problem.

The energetic variational form proposed here also opens a new door to design a positiveness preserving and energy stable numerical schemes for reaction-diffusion type equations.
For instance, such an energetic variational form will enable us to design Lagrangian-Eulerian schemes for reaction-diffusion systems by applying some recently developed methods for general diffusions \cite{junge2017fully, carrillo2018lagrangian, liu2019lagrangian, carrillo2019blob}.

\section*{Acknowledge}
The authors acknowledge the partial support of NSF (Grant DMS-1759536). We thank Prof. Hong Qian, Prof. Huaxiong Huang and Dr. Shixin Xu for suggestions and helpful discussions.

\appendix

\section{Numerical Method}
 
In the appendix, we give a detailed description of the numerical method that we used to study the reaction-diffusion system in Sec. \ref{sec:out}, which is based on the energetic variational formulation proposed in this paper.

From a numerical perspective,  it is often a challenge to construct a numerical scheme that preserves the positivity and conservation of elements for reaction-diffusion systems \cite{sandu2001positive, formaggia2011positivity}. The energetic variational formulation presented in this paper opens a new door to design a positive, energy-stable numerical schemes to reaction-diffusion type equations. 
\begin{figure}[!htb]
  \centering
 \vspace{1em}
    \begin{overpic}[width = \linewidth]{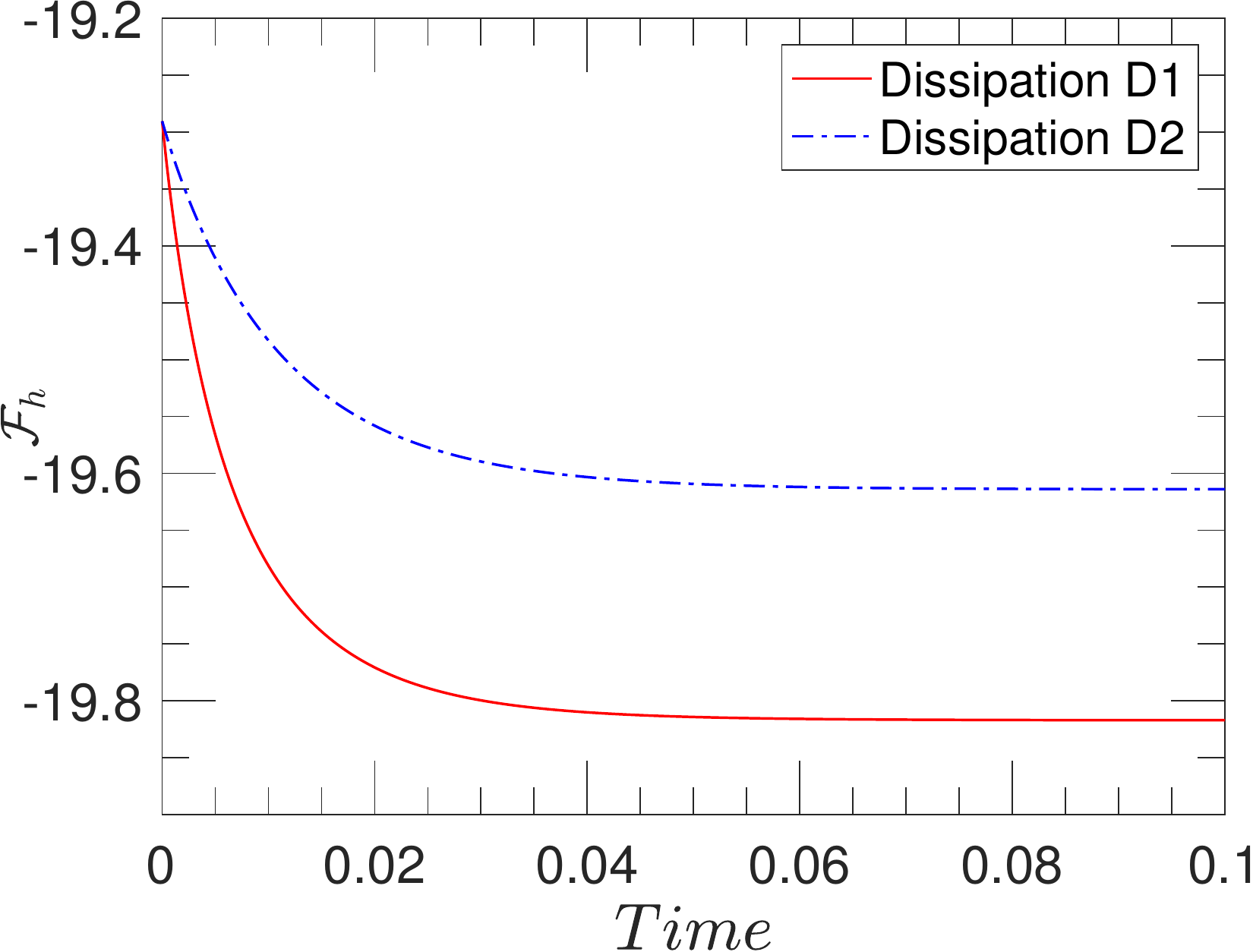}
  \end{overpic}
  \caption{Change of the discrete energy for two dissipations with respect to time in numerical simulations ($\tau = 10^{-4}$). }\label{Energy_Res}
\end{figure}

Here we only proposed a numerical scheme for the one-dimensional reaction-diffusion system considered in Sec. \ref{sec:out}.
Our numerical discretization is based a discrete energetic variational approach \cite{liu2019lagrangian, liu2020variational}, which follows the strategy of ``discretize-then-variation''.
More specifically, we can adopt a finite difference scheme on a staggered grid for the spatial discretization of $R$ and the accumulated fluxes $J_i = \int_{0}^t c_i u_i \dd t$ ($i = A, B, C$). 
Assume $[0, l]$ is the compuatational domain, let $X_j = jh$ ($j = 0, \ldots, N$) be the equidistant grid point and $X_{j+1/2} = (j + 1/2) h$ ($j = 0, \ldots, N - 1$) be the corresponding half-integer grid point, where $h = k/N$.

 Let $\mathcal{E}_N$ and $\mathcal{C}_N$ be the spaces of functions defined on  $\{ X_j ~|~ j = 0, \ldots, N \}$ and  $\{ X_{j+1/2}  ~|~ j = 0, \ldots, N - 1 \}$, respectively, We can approximate $R$ and $c_i$ in $\mathcal{E}_N$ and approximate $J_i$ in $\mathcal{C}_N$. Then the kinematic $c_i = c_i^0 + \sigma_i R + \pp_x J$ becomes
\begin{equation}\label{discrete_C}
(c_{i})_j(t) = (c_i^0)_j + \sigma_i R_j(t) + \dfrac{ (J_{i})_{j+1/2} -  (J_{i})_{j-1/2}}{h},
\end{equation}
where $i = A, B, C$. Inserting (\ref{discrete_C}) into (\ref{ED_RD}), we get the discrete energy in terms of $R_j$ and $J_{j+1/2}$. On the meantime, for the dissipation (\ref{D_1}) and (\ref{D_2}), the discrete dissipation functional can be written as
\begin{equation*}
  \begin{aligned}
    & \mathcal{D}_h =   \sum_{j=1}^N \Gamma(R_j' (t)) R_j'(t)) \\
    & \quad + \sum_{k=1}^{N-1} \left( |(J_{A})_{k+1/2}'|^2 +  |(J_{B})_{k+1/2}'|^2 +  |(J_{C})_{k+1/2}'|^2 \right). \\
   \end{aligned}
\end{equation*}
By employing a discrete energetic variational approach, we get
\begin{equation}\label{semi}
  \begin{cases}
    & \Gamma(R_j'(t)) = \left( - (\mu_A)_j - (\mu_{B})_j + (\mu_C)_j \right)  \\
    & (J_i)^{'}_{k + 1/2}  = \dfrac{ (\mu_{i})_{k+1} - (\mu_i)_k}{h}, \\
  \end{cases}
\end{equation}
where
\begin{equation}
  \begin{aligned}
    & (\mu_i)^{n+1}_j = \ln (c_i)^{n+1}_j - \ln (c_{i}^{\infty})_j, \\
  \end{aligned}
\end{equation}
$j = 0, \ldots N$ and $k = 0, \ldots N - 1$. The fully discrete scheme can be obtained by applying the implicit Euler discretization to (\ref{semi}), that is 
\begin{equation}\label{FD_scheme}
  \begin{cases}
    &  \Gamma( \dfrac{R^{n+1}_j - R^n_j}{\tau}) = \left( - \sum_{i=1}^3 \sigma_i (\mu_{i})_j^{n+1} \right),  \\  
    & \\
    & \dfrac{ (J_i)^{n+1}_{k + 1/2} - (J_i)^{n}_{k+1/2} }{\tau} = \dfrac{ (\mu_{i})_{k+1}^{n+1} - (\mu_i)_k^{n+1}}{h}/  
  \end{cases}
\end{equation}
As a numerical test, we compute our system with $c_0 = 0.25$ for dissipations (\ref{D_1}) and (\ref{D_2}). The computed discrete free energy as a function of time is showed in Fig. \ref{Energy_Res}. The simulation result indicates that our 
numerical scheme is energy stable, although a careful numerical analysis is certainly needed.


\bibliographystyle{apsrev}
\bibliography{KCR}

\end{document}